\documentclass[amsmath,amssymb,showpacs,superscriptaddress]{revtex4}

\usepackage{graphicx}
\usepackage{dcolumn}
\usepackage{bm}
\usepackage{amsmath}
\usepackage{amssymb}

\newcommand{\be}{\begin{equation}}
\newcommand{\ee}{\end{equation}}

\newcommand{\fa}{\Psi_A}

\newcommand{\fad}{\Psi_A^\dagger}

\newcommand{\lam}{\lambda}
\newcommand{\ra}{\rangle}
\newcommand{\la}{\langle}

\begin{document}

\title{One-Dimensional Impenetrable Anyons in Thermal Equilibrium. II.
Determinant Representation for the Dynamic Correlation Functions}

\author{Ovidiu I. P\^{a}\c{t}u}

\affiliation{C.N. Yang Institute for Theoretical Physics, State
University of New York at Stony Brook, Stony Brook, NY 11794-3840,
USA }
\affiliation{Institute for Space Sciences,
 Bucharest-M\u{a}gurele, R 077125, Romania}

\author{Vladimir E. Korepin}

\affiliation {C.N. Yang Institute  for Theoretical Physics, State
University of New York at Stony Brook, Stony Brook, NY 11794-3840,
USA }

\author{Dmitri V. Averin}
\affiliation{Department of Physics and Astronomy, State University
of New York at Stony Brook, Stony Brook, NY 11794-3800, USA }
\email[Electronic addresses: ]{ipatu@grad.physics.sunysb.edu ;  korepin@max2.physics.sunysb.edu ; dmitri.averin@stonybrook.edu}

\begin{abstract}

We have obtained a determinant representation for the time- and
temperature-dependent field-field correlation function of the
impenetrable Lieb-Liniger gas of anyons through direct summation of
the form factors. In the static case,  the obtained results are
shown to be equivalent to those that follow from the anyonic
generalization of Lenard's formula.

\end{abstract}
\maketitle

\section{Introduction and Statement of Results}

This is the second paper in a series that provides a comprehensive
treatment of the properties of temperature-dependent correlation
functions of one-dimensional (1D) impenetrable free anyons, based on
the methods developed for  impenetrable bosons \cite{KBI}. The
anyonic model considered in this work can be viewed as a
generalization to an arbitrary statistics parameter $\kappa$ of the
model of impenetrable bosons obtained from the Bose gas with
repulsive $\delta$-function interaction \cite{LL,KBI} in the limit
of infinitely large coupling constant (for other anyonic extensions of well known models see \cite{ AOE1,OAE1,F}).
 This model, which we call the
Lieb-Liniger gas of anyons, was formulated in this form by Kundu
\cite{Kundu}, clarified in \cite{AN,PKA} and further studied in
\cite{BGO,BG,BGH,SSC,CM,SC}. In the bosonic case, the first step in
the analysis of the correlation functions is the derivation of the
Fredholm determinant representation for these functions \cite{KS}.
With the help of the determinant representation, a classical
integrable system characterizing the correlation functions can be
constructed as in \cite{IIK1,IIKS}, leading to the short-distance
and low-density expansions of the correlators. The large-distance
asymptotics are then obtained  by the inverse scattering method for
the integrable system and the solution of the associated matrix
Riemann-Hilbert problem \cite{IIK2}. These results will be presented in future publications.
The purpose of this work is to
derive the Fredholm-determinant representation for the
temperature-dependent correlation functions in the case of anyons,
and to prove the equivalence of this representation with the anyonic
generalization of Lenard's formula \cite{PKA2}.

The results of this work can be summarized as follows. One defines
free propagators
\be e(\lam|t,x)=e^{it\lam^2-ix\lam},\ \ G(t,x)=
\int_{-\infty}^\infty e(\lam|t,x)\ d\lam\, , \ee
and the function
\be E(\mu|t,x)=\mbox{P.V.} \int_{-\infty}^\infty d\lam\ \frac{e(\lam
|t,x)}{\lam-\mu}+ e(\mu|t,x)\pi\tan\left( \frac{\pi \kappa}{2}
\right)\, , \ee
where $\mbox{ P.V.}$ denotes the Cauchy principal value. In terms of
these functions, the time- and temperature-dependent field-field
correlator of impenetrable 1D anyons is:
\be \la\fa(x_2,t_2)\fad(x_1,t_1)\ra_T=\left.e^{iht_{21}}\left(
\frac{1}{2\pi}G(t_{12},x_{12})+\frac{\partial}{\partial\alpha}\right)
\det(1+\hat{ V}_T)\right|_{\alpha=0}\, , \ee
where $x_{ab}=x_a-x_b,\ t_{ab}=t_a-t_b,\  a,b=1,2\, ,$ and
$\det(1+\hat{ V}_T)$ is the Fredholm determinant of the integral
operator with  kernel
\begin{eqnarray}
V_T(\lam,\mu) &=&\cos^2(\pi\kappa/2)\exp\left\{-\frac{i}{2}t_{12}(
\lam^2+\mu^2) +\frac{i}{2}x_{12}(\lam+\mu)\right\} \sqrt{
\vartheta(\lam) \vartheta(\mu)}\nonumber\\ & &\ \ \times
\left[\frac{E(\lam|t_{12},x_{12})-E(\mu|t_{12},x_{12})}{\pi^2(\lam-\mu)}
-\frac{\alpha}{2\pi^3}E(\lam|t_{12},x_{12})E(\mu|t_{12},x_{12})\right]\,
, \label{m1} \end{eqnarray}
which acts on an arbitrary function $f(\mu)$ as
\be \left(V_T f\right)(\lam)=\int_{-\infty}^\infty V_T(\lam,\mu)
f(\mu) \ d\mu\, . \ee
In Eq.~(\ref{m1}), $\vartheta(\lam)\equiv\vartheta(\lam,T,h)$ is the
Fermi distribution function at temperature $T$ and chemical
potential $h$
\be \vartheta(\lam,T,h)=\frac{1}{1+e^{(\lam^2-h)/T}}\, . \label{fw}
\ee
%
Introducing  integral operators $\hat K_T$ and $\hat A_T^\pm$
which act on the entire real axis and have  kernels
\be K_T(\lam,\mu)=\sqrt{\vartheta(\lam)}\frac{\sin x(\lam-\mu)}{
\lam -\mu}\sqrt{\vartheta(\mu)}\, , \ee
and
\be A_T^\pm(\lam,\mu)=\sqrt{\vartheta(\lam)}e^{\mp ix(\lam+\mu)}
\sqrt{\vartheta(\mu)}\, . \ee
we obtain the static, i.e. equal-time, correlators as
\be\label{i7} \la\fad(x)\fa(-x)\ra_{T}= \frac{1}{2\pi}\mbox{Tr
}\left[(1-\gamma\hat K_T)^{-1}\hat A_T^+\right] \det(1-\gamma\hat
K_T)|_{\gamma=(1+e^{+i\pi\kappa})/\pi}\, , \ee
and
\be\label{i9} \la\fad(-x)\fa(x)\ra_{T}= \frac{1}{2\pi}\mbox{Tr
}\left[(1-\gamma\hat K_T)^{-1}\hat A_T^-\right] \det(1-\gamma\hat
K_T)|_{\gamma=(1+e^{-i\pi\kappa})/\pi}\, . \ee
Here $\mbox{ Tr }[ f(x,y)] \equiv \int f(x,x)\ dx$, and due to the
nonconservation of parity the corrrelator $\la\fad(x)\fa(-x)\ra_{T}$
is different from $\la\fad(-x)\fa(x)\ra_{T}$.

The paper is organized as follows. Section \ref{sect2} introduces
the Lieb-Liniger gas of anyons and presents the Bethe Ansatz
eigenfunctions, Bethe equations, the ground state, and the
thermodynamics of anyons in the impenetrable limit. In Section
\ref{sect3} we compute the form factors and express the field
correlator as a Fredholm determinant. Section \ref{sect4} presents
the proof of the equivalence of Eqs.~(\ref{i7}) and (\ref{i9}) to
the anyonic version of Lenard's formula \cite{PKA2}. Some technical
details of the calculations are relegated to the Appendices.

\section{The Lieb-Liniger gas of Impenetrable Anyons}\label{sect2}

The second-quantized Hamiltonian of the Lieb-Liniger gas of 1D
anyons is
\be\label{hama} H=\int_{-L/2}^{L/2} dx \ \left( [\partial_x \fad(x)]
[\partial_x\fa(x)] +c\fad(x)\fad(x)\fa(x)\fa(x)-h\fad(x)\fa(x)
\right) \, , \ee
where $c>0$ is the coupling constant, $L$ is the length of
normalization interval, and $h$ is the chemical potential.  The
canonical Heisenberg fields
\be \fad(x,t)=e^{iHt}\fad(x)e^{-iHt}\, ,\ \ \ \fa(x,t)=e^{iHt}
\fa(x) e^{-iHt}\, , \ee
obey the anyonic equal-time commutation relations
\be\label{com1}
\fa(x_1,t)\fad(x_2,t)=e^{-i\pi\kappa\epsilon(x_1-x_2)} \fad(x_2,t)
\fa(x_1,t)+ \delta(x_1-x_2)\, , \ee
\be\label{com2}
\fad(x_1,t)\fad(x_2,t)=e^{i\pi\kappa\epsilon(x_1-x_2)} \fad(x_2,t)
\fad(x_1,t)\, , \ee
\be\label{com3}
\fa(x_1,t)\fa(x_2,t)=e^{i\pi\kappa\epsilon(x_1-x_2)}
\fa(x_2,t) \fa(x_1,t)\, , \ee
where $\kappa$ is the statistics parameter, which we assume to be
rational (this is necessary in order for  Eq. (\ref{nonvanish}) to hold), and $\epsilon(x)=x/|x|,\ \epsilon(0)=0.$ The Fock vacuum
is defined as usual by
\be \label{vacuum} \fa(x)|0\ra=0=\la 0|\fad(x)\, , \ \ \la
0|0\ra=1\, . \ee
The eigenstates $|\Psi_N\ra$ of the Hamiltonian are
\be\label{eigen}
|\Psi_N\ra=\frac{1}{\sqrt{N!}}\int_{-L/2}^{L/2}dz_1\cdots\int_{-L/2}^{L/2}dz_N\
\chi_N(z_1,\cdots,z_N|\lam_1,\cdots,\lam_N)\fad(z_N)\cdots\fad(z_1)|0\ra\,
,\ee
where  quantum-mechanical wavefunctions have the property of
anyonic exchange statistics:
\be
\chi_N(\cdots,z_i,z_{i+1},\cdots)=e^{i\pi\kappa\epsilon(z_i-z_{i+1})}
\chi_N(\cdots,z_{i+1},z_i,\cdots)\, . \label{m8} \ee
Note that the sign in front of the statistical phase in this
expression ($+i\pi \kappa$ or $-i\pi \kappa$) depends on the choice
of ordering of the creation operators in the definition of the
eigenstates (\ref{eigen}). The order of these operators adopted in
Eq.~(\ref{eigen}) (leading to the phase $+i\pi \kappa$): the
particle with the first coordinate $z_1$ created first, then $z_2$,
etc., is convenient \cite{AN} for the subsequent calculation of the
form factors.

In this paper, we limit our discussion to the case of infinitely
strong interaction, $c \rightarrow \infty$, which corresponds to
impenetrable anyons. In general, the eigenfunctions $\chi_N$ are
\cite{PKA}
\be \chi_N = \frac{e^{+i\frac{\pi\kappa}{2}
\sum_{j<k}\epsilon(z_j-z_k)}}{ \sqrt{N!\prod_{j>k}[(\lambda_j-
\lambda_k)^2 +c'^2]}} \sum_{\pi\in S_N}(-1)^{\pi}e^{i\sum_{n=1}^Nz_n
\lambda_{\pi(n)}} \prod_{j>k}[\lambda_{\pi(j)}-\lambda_{ \pi(k)}
-ic' \epsilon(z_j-z_k)] \, , \label{m0} \ee
where $c'\equiv c/\cos(\pi\kappa/2)$, and reduce for impenetrable
anyons to a simpler form:
\be \chi_N =\frac{e^{ +i\frac{\pi \kappa}{2}\sum_{j<k} \epsilon(z_j
-z_k)}}{\sqrt{N!}}\prod_{j>k} \epsilon(z_j-z_k) \sum_{\pi\in
S_N}(-1)^\pi e^{i\sum_{n=1}^N z_n\lam_{\pi(n)}} \label{anwav}\, .
\ee
Here $S_N$ is the group of permutations of $N$ elements, and
$(-1)^\pi$ is the sign of the permutation. The energy eigenvalues
\[ H|\Psi_N\ra=E|\Psi_N\ra \]
are given by the sum of effectively single-particle contributions:
\be E=\sum_{i=1}^N\varepsilon(\lam_j), \ \mbox{ with } \ \ \
\varepsilon(\lam)=\lam^2-h\,. \ee

The individual momenta $\lambda_j$ depend of the boundary conditions
imposed on the wavefunctions. In contrast to particles of integer
statistics, wavefunctions of the anyons satisfy different
quasi-periodic boundary conditions in their different arguments, the
difference resulting from the statistical phase shift $2\pi \kappa$
\cite{AN,PKA}. In general, the quasi-periodic boundary conditions
also include the external phase shift $\eta$ (we will consider $\eta$=$2\pi\times$ rational), so that the boundary
conditions on the wavefunctions (\ref{anwav}) are:
\begin{eqnarray}
\chi_N(-L/2,z_2,\cdots,z_N)& =& e^{-i \eta} \ \ \ \ \
\chi_N(L/2,z_2,\cdots,z_N) \, , \nonumber
\\ \chi_N(z_1,-L/2,\cdots,z_N) &=&e^{i(2\pi\kappa - \eta)}
\chi_N(z_1,L/2,\cdots,z_N) \, , \nonumber \\
                 & \vdots & \label{boundarycond}  \\
\chi_N(z_1,z_2,\cdots,-L/2)&=& e^{i(2\pi (N-1) \kappa -\eta)}
\chi_N(z_1,z_2\cdots,L/2) \, .  \nonumber
\end{eqnarray}
The difference in the boundary conditions for different arguments of
$\chi_N$ makes it possible, in general, to impose the condition
without the statistical phase shift on any of the arguments $z_j$.
The precise form of the Bethe equations for the momenta $\lambda_j$ in
the wavefunction (\ref{m0}) depends on specific choice of the
boundary conditions. The choice (\ref{boundarycond}), in which the
first coordinate $z_1$ does not have the statistical shift in its
boundary condition, gives rise to the Bethe equations which include the
full statistical contribution $\pi \kappa (N-1)$ to the momentum
shift of each of the anyons produced by the $N-1$ other anyons in
the system \cite{PKA}:
\be e^{i\lambda_jL}= e^{i\, \overline\eta }\prod_{k=1,k \ne
j}^N\left( \frac{ \lambda_j-\lambda_k+ic'}{\lambda_j-\lambda_k-ic'}
\right)\, ,\label{m7} \ee
where $\overline\eta=\eta- \pi \kappa (N-1)$. Similarly to the
wavefunctions, the general Bethe equations (\ref{m7}) are simplified
in the impenetrable limit $c\rightarrow \infty$:
\be e^{i\lambda_jL} = (-1)^{N-1}e^{i\, \overline \eta}\, .
\label{m2} \ee

\subsection{Structure of the Ground State}

We assume that the ground state of the gas contains $N$ anyons, and
take, for convenience, $N$ to be even, although this does not affect
our final results. We denote the momenta of the particles in the
ground state as $\mu_j$, where $j=1,\cdots,N$, and introduce the
notation $\{[...]\}$ such that
\be\label{fractionalpart} \{[x]\}=\gamma\, , \ \ \ \ \mbox{ if }
x=2\pi \times \mbox{ integer }+2\pi\gamma \, , \ \ \ \gamma\in
(-1,1) \, . \ee
The Bethe equations (\ref{m2}) give then the momenta $\mu_j$:
\be\label{GSBE}
\mu_j=\frac{2\pi}{L}\left(j-\frac{N+1}{2}\right)+\frac{2\pi\delta}{L},\
\ \ j=1,\cdots,N_0\, , \ee
where $\delta=\{[\overline\eta ]\}.$ In the thermodynamic limit
$L\rightarrow \infty,\  N\rightarrow\infty,\ N/L=D$, momenta of the
particles fill densely the Fermi sea $[-q,q],$ where $q=\sqrt{h}$ is
the Fermi momentum and the gas density is $D=q/\pi$.

\subsection{Thermodynamics}

The thermodynamics of the Lieb-Liniger anyonic gas was considered in
\cite{BG,BGH}. Similarly to the structure of the ground state, all
local thermodynamic characteristics in the case of impenetrable
anyons are equivalent to those of the free fermions. At
non-vanishing temperature $T$, the quasiparticle distribution is
given by the Fermi weight (\ref{fw}), and the density and energy are
\be D=\frac{1}{2\pi}\int_{-\infty}^\infty \vartheta(\lam,h,T)\
d\lam\, ,\ \ \ E=\frac{1}{2\pi}\int_{-\infty}^\infty\
\lam^2\vartheta(\lam,h,T)\ d\lam \, . \ee
The density increases monotonically as a function of   the chemical potential $h$.
At $T=0$,we have $D=0$ for $h\leq 0$, and $0<D<\infty$ if
$0<h<\infty$. At non-vanishing temperature, the density is zero for
$h=-\infty$ and monotonically increases with $h$ for
$-\infty<h<\infty.$

\section{Time Dependent Field-Field Correlator}\label{sect3}

In our previous paper \cite{PKA2}, we have derived the anyonic
generalization of the Lenard formula, which for impenetrable free
anyons, is an expansion of the anyonic reduced density matrices in
terms of the reduced density matrices of free fermions. In the
simplest case, the correlator
\be\label{lfr} (x_1|\rho_1^a|x_2)=\la\fad(x_2)\fa(x_1)\ra_T\,  \ee
is the first Fredholm minor of an integral operator, whose kernel is
the Fourier transform of the Fermi weight (\ref{fw}). In this
section, we obtain the time dependent generalization of this result.
Our approach will be based on the following considerations. We start
with the zero temperature field correlator
\be \la\fa(x_2,t_2)\fad(x_1,t_1)\ra_{N}
=\frac{\la\Psi(\mu_1,\cdots,\mu_{N})
|\fa(x_2,t_2)\fad(x_1,t_1)|\Psi(\mu_1,\cdots,\mu_{N})\ra} {\la \Psi(
\mu_1,\cdots,\mu_{N})|\Psi(\mu_1,\cdots,\mu_{N})\ra}\, , \ee
where the wavefunctions are taken to be normalized as
\be
{\la\Psi(\mu_1,\cdots,\mu_{N})|\Psi(\mu_1,\cdots,\mu_N)\ra}=L^N,
\label{m3} \ee
and $\mu_1,\cdots,\mu_{N}$ are the momenta in the ground state
(\ref{GSBE}). Using the resolution of identity for the Hilbert space
of $N+1$ particles
\be {\bf 1}=\sum_{\mbox{ all }
\{\lam\}_{N+1}}\frac{|\Psi(\lam_1,\cdots,\lam_{N+1})\ra\la\Psi(\lam_1,
\cdots,\lam_{N+1})|} {\la\Psi(\lam_1,\cdots,\lam_{N+1})|\Psi(\lam_1,
\cdots,\lam_{N+1})\ra}\, , \ee
where, according to (\ref{m3})
\[ \la \Psi(\lam_1,\cdots,\lam_{N+1})|\Psi(\lam_1,\cdots,
\lam_{N+1})\ra=L^{N+1}\, , \]
and the sum is over all possible solutions of the Bethe equations
with $N+1$ particles, we have
\be \la \fa (x_2,t_2) \fad (x_1,t_1) \ra_N = \frac{1}{L^{2N+1}}
\sum_{\mbox{all} \, \{ \lam \}_{ N+1}} \la \Psi_N(\{ \mu
\})|\fa(x_2,t_2)|\Psi_{N+1}(\{ \lam\}) \ra \la \Psi_{N+1}(\{ \lam
\})|\fad(x_1,t_1)|\Psi_N(\{ \mu \})\ra \, . \label{m4} \ee
Defining the form factors
\be \label{m6} F_{N+1,N}(x,t)=\la \Psi_{N+1}(\{\lam\}) |\fad(x,t)|
\Psi_N(\{ \mu\}) \ra\, ,\ \ \ F_{N+1,N}^*(x,t)=\la \Psi_{N}(\{\mu
\})|\fa(x,t) |\Psi_{N+1}(\{\lam\})\ra\ , \ee
we can rewrite Eq.~(\ref{m4}) as
\be \label{fcsum} \la \fa(x_2,t_2) \fad(x_1,t_1)\ra_N =
\frac{1}{L^{2N+1}} \sum_{\mbox{all}\, \{\lam\}_{N+1}}
F_{N+1,N}^*(x_2,t_2)F_{N+1,N}(x_1,t_1)\, . \ee
Equation (\ref{fcsum}) means that in order to find the dynamic field
correlator, we need to compute the form factors and sum over all of
them. After the summation, one can take the thermodynamic limit. In
general, such a summation of form factors is extremely difficult.
The main simplification which makes it possible to perform this
summation in the model of anyons we consider here, is the fact that,
similarly to the problem of  impenetrable bosons \cite{KBI,KS},
the local thermodynamic properties of particles with $\delta$-function
interaction are identical with those of free fermions regardless of
the actual exchange statistics. Finally, the finite-temperature
correlator can be obtained from the zero-temperature result using
the standard argument developed for the Bose gas (see, e.g.,
Appendix XIII.1 of \cite{KBI}), which is also applicable in the case
of anyons.

\subsection{Form Factors}

As a first step in carrying out the program outlined above, we
compute the form factors. In the definition (\ref{m6}) of the form
factors, the eigenstates $|\Psi_N(\{\mu\})\ra, |\Psi_{N+1}(\{
\lam\}) \ra$ have, respectively, $N$ and $N+1$ particles. Although
the set $\{\mu\}$ represents in (\ref{m6}) momenta in the ground
state of $N$ particles, our calculation below is valid also when
$|\Psi_N(\{\mu\})\ra$ is not the ground state. As before, we assume
for convenience that $N$ is even. We denote by $\{\mu_j\}$ the
momenta of the anyons in the $N$-particle eigenstate, and by
$\{\lam_j\}$ the momenta in the $N+1$ eigenstate.


Using the definition (\ref{eigen}) for the eigenstates with $N$ and
$N+1$ anyons
\[ |\Psi_N(\{\mu\})\ra=\frac{1}{\sqrt{N!}}\int d^Nz\
\chi_N(z_1,\cdots,z_N|\{\mu\})\fad(z_N)\cdots\fad(z_1)|0\ra\, , \]
\[ \la\Psi_{N+1}(\{\lam\})|=\frac{1}{\sqrt{N+1!}}\int d^{N+1}y\
\la0|\fa(y_{1})\cdots\fa(y_{N+1})\chi^*_{N+1}(y_1,\cdots,y_{N+1}|
\{\lam\}) \]
one can write the form factor as
\begin{eqnarray}
F_{N+1,N}(x,0)&=&\frac{1}{\sqrt{(N+1)!N!}}\int d^{N+1}y\ d^Nz\
\chi^*_{N+1}(y_1,\cdots,y_{N+1}|\{\lam\})\chi_N(z_1,\cdots,z_N|\{\mu\})
\cdot \nonumber \\ & & \la 0|\fa(y_{1}) \cdots \fa(y_{N+1}) \fad(x)
\fad(z_N) \cdots\fad(z_1)|0\ra\, . \label{m10}
\end{eqnarray}
A direct application of the anyonic commutation relation (\ref{com1})
and Eq.~(\ref{vacuum}) described in more details in Appendix
\ref{FFA}, reduces this expression to
\be \label{ffint} F_{N+1,N}(x,0)=\la \Psi_{N+1}|\fad(x) |\Psi_{N}\ra
=\sqrt{N+1}\int d^Nz\ \chi^*_{N+1}(z_1, \cdots,z_N,x|\{\lam\})
\chi_N(z_1, \cdots,z_N|\{\mu\})\, . \ee
An important feature of Eq.~(\ref{ffint}) is that the order of the
creation operators chosen in Eq.~(\ref{eigen}) makes the ``free''
coordinate $x$ in (\ref{ffint}) the last argument of the
wavefunction $\chi_{N+1}$. This ensures that both wavefunctions,
$\chi_{N}$ and $\chi_{N+1}$, have the same phase shifts
(\ref{boundarycond}) at the boundary of the normalization interval
in all other variables $z_j$. Since these phase shifts are canceled
in Eq.~(\ref{ffint}), the expression under the integrals over $z_j$
is periodic in each of the variable \cite{AN}. This feature is the
necessary consistency condition for the Hilbert spaces of anyon
wavefunctions with different numbers of particles, and is important
in what follows for the appropriate calculation of the form factors
(\ref{ffint}).

The sets of momenta $\{\mu_j\}$ and $\{\lam_j\}$ in the
wavefunctions $\chi_{N}$ and $\chi_{N+1}$ in (\ref{ffint}) are
determined by the Bethe equations (\ref{m2}) as
\be \mu_j=\frac{2\pi}{L}\left(m_j+\frac{1}{2}\right)+\frac{2 \pi
\delta }{L},\ \ \ \delta=\{[\eta -\pi \kappa (N-1)]\},\
j=1,\cdots,N,\ m_j\in\mathbb{Z}\, , \ee
\be \lam_j=\frac{2\pi}{L}n_j+\frac{2\pi\delta'}{L},\ \ \ \delta'
=\{[\eta -\pi \kappa N ]\},\ j=1,\cdots,N+1,\ n_j\in \mathbb{Z}\,
.\label{e30} \ee
These equations show that
\be \label{nonvanish}
\lam_j-\mu_k=\frac{2\pi}{L}\left(l-\frac{\kappa+1}{2}\right) ,\ \ \
l\in\mathbb{Z}\, , \ee
which means that $\lam_j$ and $\mu_k$ never coincide except in the
trivial case $\kappa=1$, when we have a gas of non-interacting
fermions. In all other situations, $\lam_j$ and $\mu_k$ are
different. This difference between them comes from the phase shift
due to the hard-core condition on the added particle described by
the factor $1/2$ in (\ref{nonvanish}), and the extra anyonic
statistical phase added to the anyon system together with the
particle \cite{AN}. This difference between $\lam_j$ and $\mu_k$
plays an important role in the following calculations. Using the
identity
\be e^{+i\frac{\pi\kappa}{2}\epsilon(x-y)}\epsilon(y-x)=
\cos\left(\pi\kappa/2\right)\epsilon(y-x)-i\sin\left(\pi\kappa/2
\right)\, , \ee
we can rewrite the anyonic wavefunction (\ref{anwav}) as
\be \chi_N(z_1,\cdots,z_N|\{\mu\})=\frac{\prod_{j>k}\left[\cos\left(
\pi\kappa/2\right)\epsilon(z_j-z_k)-i\sin\left(\pi\kappa/2\right)
\right]} {\sqrt{N!}}\sum_{\pi\in S_N}(-1)^\pi e^{i\sum_{n=1}^N
z_n\mu_{\pi(n)}}\, . \ee
Using this expression for both of the wavefunctions in (\ref{ffint})
we obtain
\begin{eqnarray}
F_{N+1,N}(x,0)&=&\frac{1}{N!}\sum_{\pi\in S_{N+1}}\sum_{\sigma\in
S_{N}}(-1)^{\pi+\sigma}e^{-ix\lam_{\pi(n+1)}}\nonumber\\& &\times
\int_{-L/2}^{L/2}\prod_{n=1}^Ndz_n\left[\cos\left(\pi\kappa/2\right)
\epsilon(x-z_n)+i\sin\left(\pi\kappa/2\right)\right]e^{-i\sum_{n=1}^N
z_n(\lam_{\pi(n)}-\mu_{\sigma(n)})}\, . \end{eqnarray}
Integration by parts in this equation produces the boundary terms in
the following form
\begin{equation*}
\left.\frac{e^{-iz_n(\lam_{\pi(n)}-\mu_{\sigma(n)})}}{-i(\lam_{\pi(n)}
-\mu_{\sigma(n)})}\left(\cos\left(\pi\kappa/2\right)\epsilon(x-z_n)+
i\sin\left(\pi\kappa/2\right)\right)\right|^{z_n=L/2}_{z_n=-L/2} =
\end{equation*}
\be \frac{e^{-i\frac{\pi\kappa}{2}}e^{-i\frac{L}{2}(\lam_{\pi(n)}-
\mu_{\sigma(n)})}}{i(\lam_{\pi(n)}-\mu_{\sigma(n)})} \left(1+ e^{+
i\pi\kappa}e^{iL(\lam_{\pi(n)}-\mu_{\sigma(n)})}\right) . \ee
All these terms vanish due to Eq.~(\ref{nonvanish}). Then, using the
relation
\be \frac{d\epsilon(x-z_n)}{dz_n}=-2\delta(x-z_n)\, , \ee
we obtain the following expression for the form factors
\begin{eqnarray}\label{f45}
F_{N+1,N}(x,0)&=&\frac{[2i\cos(\pi\kappa/2)]^N}{N!}\exp\left\{ix
\left[\sum_{j=1}^N\mu_j-\sum_{j=1}^{N+1}\lam_j\right]\right\}
\sum_{\pi\in S_{N+1}}\sum_{\sigma\in S_N}(-1)^{\pi+\sigma}
\prod_{j=1}^N \frac{1}{\lam_{\pi(j)}-\mu_{\sigma(j)}}\, .
\end{eqnarray}

This expression differs from the corresponding result for
impenetrable bosons \cite{KBI,KS} by the spectrum of the momenta
which now include the statistical shift, and by the overall
$[\cos(\pi\kappa/2)]^N$ factor. For $\kappa=0$, both differences
disappear, and Eq.~(\ref{f45}) reproduces, as should be, the case of
the Bose gas. We transform this equation following the corresponding
steps for bosons \cite{KBI,KS}. One can see directly that the sums
over permutations in (\ref{f45}) can be written in the form of a
determinant:
\be\label{f46} \frac{1}{N!}\sum_{\pi\in S_{N+1}}\sum_{\sigma\in
S_N}(-1)^{\pi+\sigma}\prod_{j=1}^N\frac{1}{\lam_{\pi(j)}-\mu_{
\sigma(j)}} =\left.\left(1+\frac{\partial}{\partial\alpha}\right)
\mbox{det}_N\left(M_{jk}^{\alpha}\right)\right|_{\alpha=0}\, , \ee
with
\be\label{defm}
M_{jk}^{\alpha}=\frac{1}{\lam_j-\mu_k}-\frac{\alpha}{ \lam_{N+1}-
\mu_k},\ \ j,k=1,\cdots,N\, , \ee
reducing Eq.~(\ref{f45}) to
\be F_{N+1,N}(x,0)=(2i\cos(\pi\kappa/2))^N\exp\left\{ix\left[\sum_{
j=1}^N\mu_j-\sum_{j=1}^{N+1}\lam_j\right]\right\}
\left.\left(1+\frac{\partial}{\partial\alpha}\right)
\mbox{det}_N\left(M_{jk}^{\alpha}\right)\right|_{\alpha=0}\, .
\label{f47} \ee
The determinant part of this equation can also be written as
\be \left.\left(1+\frac{\partial}{\partial\alpha}\right)
\mbox{det}_N\left(M_{jk}^{\alpha}\right)\right|_{\alpha=0}=
\sum_{\pi\in S_{N+1}}(-1)^\pi\prod_{j=1}^N\frac{1}{ \lam_{\pi(j)}
-\mu_j}\, , \ee
as one can see directly from the L.H.S. of (\ref{f46}) by noticing
that due to the permutations $\pi$ of $\lam_{j}$, all permutations
of $\mu_{j}$ give identical contributions to the sum over $\pi \in
S_{N+1}$.

Alternatively, one can introduce a fictitious momentum $\mu_{N+1}$,
and obtain the following representation \cite{M} of the form factor
in terms of this momentum:
\be F_{N+1,N}(x,0)=(2i\cos(\pi\kappa/2))^N\exp\left\{ix\left[
\sum_{j=1}^N\mu_j-\sum_{j=1}^{N+1}\lam_j\right]\right\}
\lim_{\mu_{N+1}\rightarrow\infty}\left[-\mu_{N+1}\
\mbox{det}_{N+1}\left(\frac{1}{\lambda_j-\mu_k}\right)\right]\, ,
\ee
where $\mbox{det}_{N+1}(a_{jk})$ is the determinant of the
$(N+1)\times(N+1)$ matrix with elements $a_{jk}.$ We will not be
using this representation explicitly below.

The time-dependent form factors can be obtained from the timeless
form (\ref{f47}) using the following simple relations:
\be
e^{-iHt}|\Psi_N(\{\mu\})\ra=e^{-it\sum_{j=1}^N(\mu_j^2-h)}|\Psi_N(
\{\mu\})\ra\, , \ee
and
\be \la\Psi_N(\{\lam\})|e^{iHt}=e^{it\sum_{j=1}^{N+1}(\lam_j^2-h)}
\la\Psi_N(\{\lam\})|\, . \ee
Combining the exponential factors in these expressions with those in
Eq.~(\ref{f47}), we arrive at the final result for the
time-dependent form factor:
\be\label{fffinal}
F_{N+1,N}(x,t)=(2i\cos(\pi\kappa/2))^Ne^{-iht}\left(\prod_{i=1}^{
N+1} e(\lam_i|t,x)\right)\left(\prod_{j=1}^Ne^*(\mu_j|t,x)\right)
\left.\left(1+\frac{\partial}{\partial\alpha}\right)
\mbox{det}_N\left(M_{jk}^{\alpha}\right)\right|_{\alpha=0}\, , \ee
where we have introduced the function
\be e(\lam|t,x)=e^{it\lam^2-ix\lam} \, , \label{m16} \ee
$e^*(\lam|t,x)$ is its complex conjugate, and $M_{jk}^\alpha$ is
defined in (\ref{defm}). The form factor of the annihilation
operator $\fa(x,t)$ is obtained through complex conjugation
\be\label{fffinalcc}
\la\Psi_N(\{\mu\})|\fa(x,t)|\Psi_{N+1}(\{\lam\})\ra=F^*_{N+1, N}
(x,t)\, . \ee

\subsection{Summation of the Form Factors}

Using Eqs.~(\ref{fffinal}) and (\ref{fffinalcc}), we write the field
correlator (\ref{fcsum}) as a sum over intermediate momenta
$\{\lam \}$:
\begin{eqnarray}\label{f53}
\la\fa(x_2,t_2)\fad(x_1,t_1)\ra_N&=&\sum_{\mbox{ all } \{\lam
\}_{N+1}} \frac{(2\cos(\pi\kappa/2))^{2N}}{L^{2N+1}}e^{iht_{21}}
\left(\prod_{i=1}^{N+1}e^*(\lam_i|t_{21},x_{21})\right)
\left(\prod_{j=1}^Ne(\mu_j|t_{21},x_{21})\right)\nonumber\\
& &\times\left.\left(1+\frac{\partial}{\partial\alpha}\right)
\mbox{det}_N\left(M_{jk}^{\alpha}\right)\right|_{\alpha=0}
\left.\left(1+\frac{\partial}{\partial\beta}\right)
\mbox{det}_N\left(M_{jk}^{\beta}\right)\right|_{\beta=0}\, ,
\end{eqnarray}
with the notations $x_{ab}=x_a-x_b,\ t_{ab}=t_a-t_b,\ a,b=1,2$. The
matrix $M_{jk}^\beta$ here is the same as (\ref{defm}) with $\alpha$
replaced by $\beta$. As was mentioned above, modulo the
$[\cos(\pi\kappa/2)]^{2N}$ factors and the spectrum of momenta,
Eq.~(\ref{f53}) is identical with the expression for the bosonic
field correlators \cite{KBI,KS}. This means that the summation
process over $\{\lam \}$ is very similar, and we just sketch the
derivation here. Since we sum over all momenta $\{\lam \}$,
individual momenta $\lambda_j$ are equivalent up to permutation.
This means that one of the two permutations of $\{\lam_j\}$ involved
in the definition of the two determinants in (\ref{f53}) produces
coinciding terms, so that under the sum over $\{\lam_j\}$, one can
replace one of the determinants, e.g., the second one, with
\be (N+1)!\prod_{j=1}^N\frac{1}{\lam_j-\mu_j}\, , \ee
obtaining
\begin{eqnarray}
\lefteqn {\la\fa(x_2,t_2)\fad(x_1,t_1)\ra_N} \nonumber \\ &=&
e^{iht_{21}}\left(\prod_{j=1}^Ne(\mu_j|t_{21},x_{21})\right)
\frac{1}{L}\left(\frac{2\cos(\pi\kappa/2)}{L}\right)^{2N}(N+1)!
 \nonumber\\ & \times & \sum_{\mbox{all }
\{\lam\}_{N+1}}\left(e^*(\lam_{N+1}|t_{21},x_{21})+
\frac{\partial}{\partial\alpha}\right)
\mbox{det}_N\left.\left(\frac{e^*(\lam_j|t_{21},x_{21})}{(\lam_j- \mu_k
)(\lam_j-\mu_j)} -\alpha\frac{e^*(\lam_j|t_{21},x_{21})}{(\lam_j
-\mu_j)} \frac{e^*(\lam_{N+1}|t_{21},x_{21})}{(\lam_{N+1}-\mu_j)}
\right)\right|_{\alpha=0}. \label{f55}
\end{eqnarray}
The summation over the momenta $\{\lam_j\}$ can be done then
independently  over each $\lam_j$ inside the determinant. Also, we
transfer the factors $e(\mu_j|t_{21},x_{21})$ in (\ref{f55}) into
the determinant splitting them between the rows and columns, and use
the formula
\be \frac{1}{(\lam_j-\mu_k)}\frac{1}{(\lam_j-\mu_j)}=\left(\frac{1}{
\lam_j-\mu_j}-\frac{1}{\lam_j-\mu_k}\right)\frac{1}{\mu_j-\mu_k}\, .
\ee
This gives the correlator as
\begin{eqnarray}\label{e60}
\lefteqn{\la\fa(x_2,t_2)\fad(x_1,t_1)\ra_N=e^{iht_{21}}\left(
\frac{1}{2 \pi}G_L(t_{12},x_{12})+\frac{\partial}{\partial\alpha}
\right)} \nonumber\\ & &\times\ \mbox{det}_N \left[ \delta_{jk}
\tilde E_L(\mu_k|t_{12}, x_{12})e(\mu_j|t_{21},x_{21})
+e(\mu_j|t_2,x_2)e^*(\mu_k|t_1,x_1) \cos^2(\pi\kappa/2) \right.
\nonumber\\ & &\left. \left. \times \left( \frac{2(1-\delta_{jk})
}{\pi L(\mu_j-\mu_k)} (E_L(\mu_j|t_{12} ,x_{12})
-E_L(\mu_k,t_{12},x_{12}) ) -\frac{\alpha}{L\pi^2}
E_L(\mu_j|t_{12},x_{12})E_L(\mu_k|t_{12}, x_{12})\right)
\right]\right|_{\alpha=0}\, , \end{eqnarray}
where we have defined the functions
\be\label{defg}
\frac{1}{2\pi}G_L(t,x)=\frac{1}{L}\sum_{\lam}e(\lam|t,x)\, , \ee
\be\label{defe}
\frac{1}{2\pi}E_L(\mu_k|t,x)=\frac{1}{L}\sum_{\lam}\frac{e( \lam|
t,x) }{\lam-\mu_k}\, , \ee
\be\label{defte} \tilde E_L(\mu_k|t,x)= \frac{4\cos(\pi\kappa/2
)^2}{L^2}\sum_\lam\frac{e(\lam|t,x)}{(\lam- \mu_k)^2}\, , \ee
and
$\lam=\frac{2\pi}{L}(\mathbb{Z}+\delta')$ -- see (\ref{e30}).
Formula (\ref{e60}) is the final expression for the field correlator
in the ground state of $N$ anyons on a finite interval with
quasi-periodic boundary conditions.

\subsection{Thermodynamic Limit}

In order to obtain the correlator in the thermodynamic limit, we
need to compute the large-$L$ limit of the functions (\ref{defg}),
(\ref{defe}), and (\ref{defte}). This is done in Appendix \ref{TL}
with the results
\be G(t,x)\equiv\lim_{L\rightarrow\infty}G_L(t,x)=
\int_{-\infty}^\infty e(\lam|t,x)\ d\lam\, , \label{m15} \ee
\be E(\mu_k|t,x)\equiv\lim_{L\rightarrow\infty}E_L(\mu_k|t,x)=
\mbox{P.V.} \int_{-\infty}^\infty d\lam\ \frac{e(\lam|t,x)}{\lam-
\mu_k} + e(\mu_k|t,x)\pi\tan\left(\frac{\pi\kappa}{2}\right)\, ,
\ee
\be\label{tltildee} \tilde E(\mu_k|t,x)\equiv \lim_{L\rightarrow
\infty} \tilde E_L(\mu_k|t,x)= e(\mu_k|t,x)+\frac{2\cos^2
(\pi\kappa/2) }{\pi L}\frac{\partial}{\partial \mu_k}E(\mu_k|t,x)\,
. \ee
In the thermodynamic limit $L,N\rightarrow\infty$ with $D=N/L$
constant, the anyon momenta fill densely the Fermi interval
$[-q,q]$, where $q=\sqrt{h}$ and $D=q/\pi$. In this case,
the determinant in the correlator (\ref{e60}) can be understood as the
Fredholm determinant of an integral operator. Indeed, for an
arbitrary integral operator $\hat V$, whose action on a function
$f(\lam)$ is defined by
\be (\hat V f)(\lam)=\int_{a}^{b}V(\lam,\mu)f(\mu)\ d\mu\, , \ee
the associated Fredholm determinant is (see, e.g., \cite{text})
\be \det(1+\hat V)=\lim_{n\rightarrow\infty} \left|
\begin{array}{cccc}
1+\xi V (\lam_1,\lam_1)&\xi V(\lam_1,\lam_2)&\cdots&\xi V(\lam_1,
\lam_n)\\ \xi V (\lam_2,\lam_1)&1+\xi V(\lam_2,\lam_2)&\cdots&\xi
V(\lam_2,\lam_n)\\ \vdots& \vdots& \ddots& \vdots\\
\xi V (\lam_n,\lam_1)&\xi V(\lam_n,\lam_2)&\cdots& 1+\xi
V(\lam_n,\lam_n) \end{array}\right|\, , \ee
where $\xi=(b-a)/n,\ \lam_p-\lam_{p-1}=\xi$ and $\lam_0=a,\
\lam_n=b.$ One can see directly that, in the thermodynamic limit,
the determinant part of Eq.~(\ref{e60}) has the same structure with
$N$ momenta $\mu_j$ separated by $\xi=2\pi/L$ filling the Fermi
interval $[-q,q]$. This means that the correlator can be expressed
as
\be \la\fa(x_2,t_2)\fad(x_1,t_1)\ra= \left.e^{iht_{21}} \left(
\frac{1}{2\pi}G(t_{12},x_{12})+\frac{\partial}{\partial\alpha}\right)
\det(1+ \hat{\tilde V}_0)\right|_{\alpha=0}\, , \ee
where $\hat{\tilde V}_0$ acts on an arbitrary function $f(\lam)$ as
\be (\hat{\tilde V}_0 f)(\lam)=\int_{-q}^{q}\tilde V_0(\lam,\mu)
f(\mu)\ d\mu\, , \ee
and
\be \tilde V_0(\lam,\mu)=\cos^2 (\pi\kappa/2) e(\lam|t_2,x_2)
e^*(\mu|t_1,x_1) \left[\frac{E(\lam|t_{12},x_{12})- E(\mu|t_{12},
x_{12})}{\pi^2(\lam-\mu)}-\frac{\alpha}{2\pi^3}E(\lam|t_{12},
x_{12})E(\mu|t_{12},x_{12})\right]\, . \label{m13} \ee
Performing the unitary transformation
\be V_0(\lam,\mu)=\exp\left\{-i\frac{(t_1+t_2)}{2}(\lam^2-\mu^2)
+i\frac{(x_1+x_2)}{2}(\lam-\mu)\right\}\tilde V_0(\lam,\mu)\, , \ee
with the property
\be \det(1+\hat{\tilde V}_0)=\det(1+\hat{V}_0)\, , \ee
we transform the kernel $\tilde V_0(\lam,\mu)$ (\ref{m13}) into the
symmetric form:
\begin{eqnarray}\label{v0final}
V_0(\lam,\mu)&=&\cos^2(\pi\kappa/2)\exp\left\{-\frac{i}{2} t_{12}(
\lam^2+\mu^2)+\frac{i}{2}x_{12}(\lam+\mu)\right\}\nonumber \\
& &\ \ \times \left[\frac{E(\lam|t_{12},x_{12})-E(\mu|t_{12},
x_{12})}{\pi^2(\lam-\mu)}-\frac{\alpha}{2\pi^3}E(\lam|t_{12},x_{12}
)E(\mu|t_{12},x_{12})\right]\, . \end{eqnarray}
Two observations are in order. First, one can check that the second
term in (\ref{tltildee}) is obtained from the first term in the
square bracket of (\ref{v0final}) in the limit $\lam\rightarrow
\mu$. Second, in the limit $\kappa\rightarrow 0$,
Eq.~(\ref{v0final}) reproduces the known result \cite{KBI,KS} for
impenetrable bosons.

In the static case ($t_1=t_2$), which is discussed in the next
Section, the kernel (\ref{v0final}) can be simplified further. One
needs to distinguish two cases.
\begin{itemize}
\item $x_1>x_2$. In this case,
\be E(\lam|0,x_{12})=-i\pi e^{-ix_{12}\lam}[1+i\tan(\pi\kappa/2)]\,
, \ee
and the kernel (\ref{v0final}) becomes
\be\label{v0pos}
V_0^+(\lam,\mu)=-\frac{(1+e^{+i\pi\kappa})}{\pi}\left(\frac{\sin
(x_{12}(\lam-\mu)/2)}{\lam-\mu}\right)
+\frac{\alpha}{2\pi}e^{+i\pi\kappa}\exp\left\{-i\frac{x_{12}}{2}(
\lam+\mu) \right \} \, . \ee
\item $x_1<x_2$. In this case,
\be E(\lam|0,x_{12})=i\pi e^{-ix_{12}\lam}[1-i\tan(\pi\kappa/2)]\, ,
\ee and
\be\label{v0neg}
V_0^-(\lam,\mu)=\frac{(1+e^{-i\pi\kappa})}{\pi}\left(\frac{ \sin(
x_{12} (\lam-\mu)/2)}{\lam-\mu}\right) +\frac{\alpha}{2\pi}
e^{-i\pi\kappa} \exp\left\{-i\frac{x_{12}}{2}(\lam+\mu)\right\}\, .
\ee
\end{itemize}

We now extend the discussion to the situation of  non-vanishing
temperature $T$. The temperature-dependent field correlator is
defined as
\be \la \fa(x_2,t_2) \fad(x_1,t_2)\ra_T =\frac{\mbox{Tr} \left(
e^{-H/T} \fa(x_2,t_2)\fad(x_1,t_1)\right)}{\mbox{Tr } e^{-H/T}}\, .
\ee
According to the well-known argument developed for the Bose gas
\cite{KBI}, this correlator can be found as the mean value over any
one of the ``typical'' eigenfunctions $\Omega_T$ of the Hamiltonian
which characterizes the given state of thermal equilibrium:
\be \frac{\la\Omega_T|\fa(x_2,t_2)\fad(x_1,t_2)|\Omega_T\ra}{\la
\Omega_T|\Omega_T\ra} \, . \label{m14} \ee
This argument depends only on the general saddle-point approximation
in the description of the state of equilibrium, and also holds in
the case of anyons. The further computation of the field correlator
based on Eq.~(\ref{m14}) is similar to the zero-temperature case,
the main difference being the change of the measure of integration:
\be \int_{-q}^q\ d\lam \;\;\; \rightarrow \;\;\;
\int_{-\infty}^\infty\ d\lam\ \vartheta(\lam,T,h)\ \ \ \mbox{with} \
\ \vartheta(\lam,T,h)=\frac{1}{1+e^{(\lam^2-h)/T}}\, . \ee
The final result for the temperature-dependent correlator is then
\be \la \fa(x_2,t_2)\fad(x_1,t_1)\ra_T =\left. e^{iht_{21}}
\left(\frac{1}{2\pi}G(t_{12},x_{12})+\frac{\partial}{\partial
\alpha}\right) \det(1+\hat{V}_T)\right|_{\alpha=0}\, , \ee
where the kernel of the integral operator $\hat{V}_T$ is
\begin{eqnarray}\label{genkern}
V_T(\lam,\mu)&=&\sqrt{\vartheta(\lam)}V_0(\lam,\mu)\sqrt{
\vartheta(\mu)}\, ,\nonumber\\ &=&\cos^2(\pi\kappa/2) \exp \left\{
-\frac{i}{2}t_{12}(\lam^2+\mu^2)+\frac{i}{2}x_{12}(\lam+\mu)
\right\} \sqrt{\vartheta(\lam)\vartheta(\mu)}\nonumber\\
& & \times \left[\frac{E(\lam|t_{12},x_{12}) -E(\mu|t_{12}, x_{12})
}{\pi^2(\lam-\mu)} -\frac{\alpha}{2\pi^3} E(\lam|t_{12},x_{12})
E(\mu|t_{12},x_{12})\right]\, , \end{eqnarray}
and the operator acts on an arbitrary function $f(\mu)$ as
\be \left(V_T f\right)(\lam)=\int_{-\infty}^\infty
V_T(\lam,\mu)f(\mu)\ d\mu\, . \ee

\section{Equivalence with Lenard Formula}\label{sect4}

In the earlier paper \cite{PKA2}, we obtained the anyonic
generalization of the Lenard formula for the equal-time field
correlator or, equivalently, reduced density matrices of anyons. In
the case of the first reduced density matrix, the anyonic Lenard
formula reads
\be \label{eq11} (x|\rho_1^a|x')_\pm=\frac{1}{\pi}\det\left.\left(
1-\gamma\hat \theta_T^\pm\left|\begin{array}{c} x\\ x'
\end{array}\right.\right)\right|_{\gamma=(1+e^{\pm i \pi \kappa})/
\pi}\, , \ee
where the kernel of the integral operators $\hat \theta^\pm_T$ is
\be\label{eq97} \theta_T(\xi-\eta)=\frac{1}{2}\int_{-\infty}^\infty
d\lam\ \frac{e^{i(\xi-\eta)\lam}}{1+e^{(\lam^2-h)/T}}\, , \ee
and their action on an arbitrary function is defined as
\be (\hat \theta^\pm_T f)(\xi)= \int_{I_\pm}
\theta_T(\xi-\eta)f(\eta)\ d\eta\, . \label{eq98} \ee
In these expressions, the plus sign refers to the situation when
$x'>x$ and $I_+ =[x,x']$, and the minus sign -- to the situation
when $x'<x$ and $I_- =[x',x]$. The resolvent kernels associated with the kernel
$\theta_T(x,y)$ acting on the intervals $I_\pm$ are denoted by $\varrho_T^\pm(\xi,\eta)$ and satisfy
the equations:
\be \varrho_T^\pm(\xi,\eta)-\frac{(1+e^{\pm i\pi\kappa})}{ \pi}
\int_{ I_{\pm}}\theta_T(\xi-\xi')\varrho_T^\pm(\xi',\eta) d\xi' =
\theta_T(\xi-\eta)\, . \ee
One can rewrite Eq.~(\ref{eq11}) in terms of the resolvent kernel
$\varrho_T$ and the field correlator as \cite{PKA2}
\be\label{eq90} \la \fad(x')\fa(x)\ra_{T,\pm} =\frac{1}{\pi}
\varrho_T^\pm(x',x) \left.\det\left(1-\gamma \hat \theta_T^\pm
\right) \right|_{\gamma=(1+e^{\pm i \pi \kappa})/\pi} , \ee
where again, the plus sign refers to the case $x'>x$ and the minus
sign -- to $x<x'$. Next, we show that Eq.~(\ref{eq90}) is reproduced
by the results obtained in the previous section when they are
specialized to the equal-time correlators. We treat the two cases,
$x'>x$ and $x'<x$, separately.

\subsection{The static correlator $\la\fa(-x)\fad(x)\ra_T$}

Equations (\ref{m16}) and (\ref{m15}) show that in the static case
\be \frac{1}{2\pi}G(0,x)=\delta(x)\, . \ee
Using this relation and Eqs.~(\ref{v0pos}) and (\ref{genkern}), we
see that the equal-time field correlator can be written as
\be\label{eq92}
\la\fa(-x)\fad(x)\ra_T=\left(\delta(2x)+\frac{\partial}{ \partial
\alpha}\right) \det\left.\left(1-\frac{(1+e^{i\pi\kappa})}{\pi} \hat
K_T +\alpha\frac{e^{i\pi\kappa}}{2\pi}\hat A_{T}^+ \right)
\right|_{\alpha=0}\, , \ee
where $\hat K_T$ and $\hat A_T^+$ are the integral operators acting
on the real axis and defined by kernels
\be\label{kt} K_T(\lam,\mu)=\sqrt{\vartheta(\lam)}\frac{\sin
x(\lam-\mu)}{\lam-\mu}\sqrt{\vartheta(\mu)}\, , \ee and \be
A_T^+(\lam,\mu)=\sqrt{\vartheta(\lam)}e^{-ix(\lam+\mu)}\sqrt{
\vartheta(\mu)}\, . \ee
At zero temperature, both operators act on the interval $[-q,q]$ and
their kernels are
\be K(\lam,\mu)=\frac{\sin x(\lam-\mu)}{\lam-\mu}\, ,\ \ \ \ \ \ \
A^+(\lam,\mu)=e^{-ix(\lam+\mu)}\, . \ee

The commutation relation (\ref{com1}) shows that
\be \la \fa(-x)\fad(x)\ra_T=e^{i\pi\kappa}\la \fad(x) \fa(-x) \ra_T
+ \delta(2x) \, . \ee
This means that in order to prove the equivalence with Lenard
formula, we have to show that
\be G^+(\kappa,x,T)\equiv\frac{\partial}{\partial\alpha} \det \left.
\left(1-\frac{(1+e^{i\pi\kappa})}{\pi}\hat K_T + \alpha \frac{ e^{i
\pi \kappa}}{2\pi}\hat A_{T}^+ \right) \right|_{\alpha =0}= e^{i \pi
\kappa}\la \fad(x)\fa(-x)\ra_T\, , \ee
where $\la \fad(x) \fa(-x) \ra_T$ is given by (\ref{eq90}). For a
general integral operator with kernel $V$, one of the useful
expressions for the Fredholm determinant is
\[ \ln \, \det(1-\gamma\hat V)=-\sum_{n=1}^\infty \frac{\gamma^n
}{n}\mbox{Tr }V^n\, . \]
Making use of this formula, we obtain
\be \label{eq99} G^+(\kappa,x,T)= \frac{e^{i\pi\kappa}}{ 2\pi }
\mbox{Tr }\left[(1-\gamma\hat K_T)^{-1}\hat A_T^+\right] \det(1-
\gamma\hat K_T)|_{\gamma=(1+e^{i\pi\kappa})/\pi}\, . \ee
Denoting as $f_-^+(\lam)$ the solution of the integral equation
\be f_-^+(\lam)-\frac{(1+e^{i\pi\kappa})}{\pi}\int_{-\infty}^\infty
K_T(\lam,\mu)f_-^+(\mu)\ d\mu=\sqrt{\vartheta(\lam)}e^{-ix\lam}\, ,
\label{m21} \ee
we can rewrite (\ref{eq99}) as
\be G^+(\kappa,x,T)=\frac{e^{i\pi\kappa}}{2\pi}\int_{-\infty}^\infty
e^{-ix\lam}f_-^+(\lam)\sqrt{\vartheta(\lam)}d\lam\ \ \det(1-\gamma\hat
K_T)|_{\gamma=(1+e^{i\pi\kappa})/\pi}\, . \ee

We will show now that
\be \label{eq102} \det(1-\gamma\hat K_T)= \det\left(1-\gamma \hat
\theta_T^+\right) \, , \ee
where the operator $\hat \theta_T$ is described by Eqs.~(\ref{eq97})
and (\ref{eq98}), and $\gamma = (1+e^{ i \pi \kappa})/\pi $. Direct
and inverse Fourier transforms of a function $g$ can be defined to
include as integration measure $\sqrt{\vartheta(\lam)}$:
\be \tilde g(\lam) =\frac{1}{2\pi\sqrt{\vartheta(\lam)}}
\int_{-\infty}^\infty d\xi\ e^{i\lam\xi}g(\xi)\, , \;\;\;\;\;
g(\xi)=\int_{-\infty}^\infty d\lam\ \sqrt{\vartheta(\lam)}
e^{-i\lam\xi}\tilde g(\lam)\, . \ee
With this definition, taking the Fourier transform of the integral
equation
\be g(\xi)-\gamma \int_{-x}^x \theta_T(\xi-\xi')g(\xi')\
d\xi'=G(\xi)\, , \ee
we obtain
\be \tilde g(\lam)-\gamma \int_{- \infty }^\infty
K_T(\lam-\mu)\tilde g(\mu)\ d\mu=\tilde G(\lam)\, . \ee
Coincidence of the two equations implies the equality (\ref{eq102})
of the determinants.

The final step in proving the equivalence of Eqs.~(\ref{eq90}) and
(\ref{eq92}) is to show that
\be\label{eq106} \varrho_T^+(x,-x)=\frac{1}{2}\int_{-\infty}^\infty
e^{-ix\lam}f_-^+(\lam)\sqrt{\vartheta(\lam)}d\lam\, . \ee

The Fourier transform of the equation defining the resolvent kernel
$\varrho_T$
\be \varrho^+_T(\xi,-x)-\frac{(1+e^{ i \pi \kappa})}{\pi}
\int_{-x}^x \theta_T(\xi-\xi')\varrho^+_T(\xi',-x)\ d\xi'
=\theta_T(\xi+x)\, , \ee
gives
\be \tilde\varrho_T^+(\lam,-x) -\frac{(1+e^{i\pi\kappa})}{\pi}
\int_{-\infty}^\infty K_T(\lam-\mu)\tilde\varrho_T^+(\mu,-x)\ d\mu
=\frac{1}{2} e^{-ix\lam}\sqrt{\vartheta(\lam)}\, . \ee
Comparison of this equation with the definition of $f_-^+(\lam)$
(\ref{m21}) shows that
\be\label{eq109} \tilde\varrho_T^+(\lam,-x)=\frac{1}{2}f_-^+(\lam)\, .
\ee
Taking the inverse Fourier transform of (\ref{eq109}) proves (\ref{eq106}).
Thus, we have shown that for $x'>x$, the Lenard formula (\ref{eq90})
is equivalent with the result (\ref{eq92}) for the static field
correlator that follows from the direct summation of the form
factors.

\subsection{The static correlator $\la\fa(x)\fad(-x)\ra_T$}

In this  case, the proof of the equivalence of the two approaches is
very similar to what was just discussed for $x'>x$. Equations
(\ref{v0neg}) and (\ref{genkern}) show that the static field
correlator is
\be \la\fa(x)\fad(-x)\ra_T=\left(\delta(2x)+\frac{\partial}{
\partial \alpha}\right) \det\left. \left(1- \frac{(1+e^{-i \pi
\kappa})}{\pi} \hat K_T+\alpha\frac{e^{-i\pi\kappa}}{2\pi}\hat
A_{T}^-\right) \right|_{\alpha=0}\, , \ee
where $\hat K_T$ is given by (\ref{kt}) and
\be A_T^-(\lam,\mu)=\sqrt{\vartheta(\lam)}e^{ix (\lam+\mu)} \sqrt{
\vartheta(\mu)}\, . \ee
From the commutation relation (\ref{com1}) we see that
\be \la\fa(x)\fad(-x)\ra_T=e^{-i\pi\kappa}\la \fad(-x)\fa(x) \ra_T
+\delta(2x)\, , \ee
so we have to show that
\be G^-(\kappa,x,T)\equiv\frac{\partial}{\partial\alpha}\det\left.
\left(1-\frac{(1+e^{-i\pi\kappa})}{\pi}\hat K_T +\alpha \frac{ e^{-i
\pi\kappa} }{2\pi}\hat A_{T}^-\right) \right|_{\alpha=0} =e^{-i\pi
\kappa}\la \fad(-x)\fa(x)\ra_T\, . \ee
where $\la \fad(-x)\fa(x)\ra_T$ is given by Eq.~(\ref{eq90}).
Similarly to the discussion in the  previous section, we can rewrite $G^-$ as
\be
G^-(\kappa,x,T)=\frac{e^{-i\pi\kappa}}{2\pi}\int_{-\infty}^\infty
e^{+ix\lam}f_+^-(\lam)\sqrt{\vartheta(\lam)}d\lam\ \ \det(1-\gamma\hat
K_T)|_{\gamma=(1+e^{-i\pi\kappa})/\pi}\, , \ee
where $f_+^-(\lam)$ is the solution of the integral equation
\be f_+^-(\lam)-\frac{(1+e^{-i\pi\kappa})}{\pi}\int_{-\infty}^\infty
K_T(\lam,\mu)f_+^-(\mu)\ d\mu=\sqrt{\vartheta(\lam)}e^{+ix\lam}\, .
\ee
The equality of the Fredholm determinants of the operators $\hat
K_T$ and $\hat \theta_T$ was shown in the previous Section, so it
remains to prove that
\be\label{eq116} \varrho_T^-(-x,x)=\frac{1}{2}\int_{-\infty}^\infty
e^{+ix\lam}f_+^-(\lam)\sqrt{\vartheta(\lam)}d\lam\, . \ee
Again, taking the Fourier transform of
\be \varrho^-_T(\xi,x)-\frac{(1+e^{ -i \pi \kappa})}{\pi}\int_{-x}^x
\theta_T(\xi-\xi')\varrho^+_T(\xi',x)\ d\xi' =\theta_T(\xi-x)\, ,
\ee
we obtain
\be \tilde\varrho_T^-(\lam,x)-\frac{(1+e^{-i\pi\kappa})}{\pi}\int_{
-\infty}^\infty K_T(\lam-\mu)\tilde\varrho_T^-(\mu,x)\ d\mu
=\frac{1}{2} e^{+ix\lam}\sqrt{\vartheta(\lam)}\, , \ee
which shows that
\be\label{eq119} \tilde\varrho_T^-(\lam,x)=\frac{1}{2}f_+^-(\lam)\, .
\ee
The inverse Fourier transform of (\ref{eq119}) gives the correct
result (\ref{eq116}).

\section{Conclusions}

In summary, we have obtained the time- and temperature-dependent
correlation functions of fields for impenetrable 1D anyons as
Fredholm determinants. The Fourier transform of the corresponding
integral equations proves the equivalence of our approach with the
anyonic Lenard formula derived previously (Eq.~57 of \cite{PKA2})
for the one-particle reduced density matrix of anyons. The same
technique can be used to obtain the multi-point correlation
functions from the Lenard formula for $n$-particle reduced density
matrices (Eq.~56 of \cite{PKA2}). The next step in the exact
calculation of the anyonic correlation functions is to use the
determinant representation derived in this work to obtain a classical
integrable system of nonlinear differential equations characterizing
these functions. These equations should make it possible to
construct the short-distance and low-density expansions for the
correlators. This will be addressed in a future publication.

\acknowledgments

This work was supported in part by the NSF grants DMS-0503712,
DMR-0325551 and 0653342.


\appendix


\section{Anyonic Form Factors}\label{FFA}

In this appendix, we prove Eq.~(\ref{ffint}). Consider first the
simple example of the form factor $F_{3,2}$:
\be\label{f32} F_{3,2}(x)=\frac{1}{2\sqrt{3}}\int d^3y\  d^2z\
\chi_3^*(y_1,y_2,y_3)\chi_2(z_1,z_2) \la
0|\fa(y_1)\fa(y_2)\fa(y_3)\fad(x)\fad(z_2)\fad(z_1)|0\ra\, . \ee
If one defines
\be A=\la 0|\fa(y_1)\fa(y_2)\fa(y_3)\fad(x)\fad(z_2)\fad(z_1) |0\ra
\, , \ee
then successive applications of the commutation relation
(\ref{com1}) followed by the Eq.~(\ref{vacuum}) give
\begin{eqnarray}
A&=&\la 0|\fa(y_1)\fa(y_2)\left[\fad(x)\fa(y_3)e^{-i\pi\kappa \epsilon
(y_3-x)}+\delta(y_3-x)\right]\fad(z_2)\fad(z_1)|0\ra \nonumber\\
&=&\la 0|\fa(y_1)\fa(y_2)\fad(x)\left[\fad(z_2)\fa(y_3)e^{-i\pi
\kappa \epsilon(y_3-z_2)}+\delta(y_3-z_2)\right]\fad(z_1)|0\ra
e^{-i\pi\kappa\epsilon(y_3-x)}\nonumber\\ & &+\la 0|\fa(y_1)
\fa(y_2) \fad(z_2)\fad(z_1)|0\ra\delta(y_3-x) \nonumber\\
&=&\underbrace{\la 0|\fa(y_1)\fa(y_2)\fad(x)\fad(z_2)|0\ra\delta
(y_3-z_1) e^{-i\pi\kappa[ \epsilon(y_3-z_2)+\epsilon(y_3-x)]}}_{{\bf
(a)}}\nonumber\\
& &+\underbrace{\la 0|\fa(y_1)\fa(y_2)\fad(x)\fad(z_1)|0\ra\delta(
y_3-z_2)e^{-i\pi\kappa\epsilon(y_3-x)}}_{{\bf (b)}}\nonumber\\
& &+\underbrace{\la 0|\fa(y_1)\fa(y_2)\fad(z_2)\fad(z_1)|0
\ra\delta(y_3-x)}_{{\bf (c)}}\, .\end{eqnarray}
Performing similar transformations, we obtain
\begin{eqnarray}
\mathbf{a}&=&\delta(y_1-x)\delta(y_2-z_2)\delta(y_3-z_1)e^{-i\pi
\kappa[\epsilon(y_2-x)+\epsilon(y_3-z_2)+\epsilon(y_3-x)]}\nonumber\\
   & &+\delta(y_1-z_2)\delta(y_2-x)\delta(y_3-z_1)e^{-i\pi\kappa
   [\epsilon(y_3-z_2)+\epsilon(y_3-x)]}\, ,\\
\mathbf{b}&=&\delta(y_1-x)\delta(y_2-z_1)\delta(y_3-z_2)e^{-i\pi
\kappa[\epsilon(y_2-x)+\epsilon(y_3-x)]}\nonumber\\
    & &+\delta(y_1-z_1)\delta(y_2-x)\delta(y_3-z_2)e^{-i\pi\kappa
    \epsilon(y_3-x)}\, ,\\ \mathbf{c}&=&\delta(y_1-z_2)\delta(y_2-z_1)
    \delta(y_3-x)e^{-i\pi\kappa\epsilon(y_2-z_2)}+\delta(y_1-z_1)\delta
    (y_2-z_2)\delta(y_3-z_3)\, . \end{eqnarray}
Substituting $A=\mathbf{a}+\mathbf{b}+ \mathbf{c}$ into (\ref{f32}),
we have for the form factor
\begin{eqnarray}\label{intff}
F_{3,2}(x)&=&\frac{1}{2\sqrt{3}}\int d^2z\ \left\{\chi_3^*
(x,z_2,z_1) \chi_2(z_1,z_2) e^{-i\pi\kappa\left[\epsilon(z_2-x)+
\epsilon(z_1-z_2)+\epsilon(z_1-x)\right]}\right.\nonumber\\
& &+\chi^*_3(z_2,x,z_1)\chi_2(z_1,z_2)e^{-i\pi\kappa\left[ \epsilon
(z_1-z_2)+\epsilon(z_1-x)\right]} +\chi^*_3(x,z_1,z_2)\chi_2(z_1,z_2)
e^{-i\pi\kappa\left[\epsilon(z_1-x)+\epsilon(z_2-x)\right]}\nonumber\\
& &\left.+\chi^*_3(z_1,x,z_2)\chi_2(z_1,z_2)e^{-i\pi\kappa
(z_2-x)}+\chi^*_3(z_2,z_1,x)\chi_2(z_1,z_2)e^{-i\pi\kappa\epsilon
(z_1-z_2)} \right. \nonumber\\ & & \left.
+\chi^*_3(z_1,z_2,x)\chi_2(z_1,z_2)\right\} . \end{eqnarray}
Using the anyonic property (\ref{m8}) of the wavefunctions, and its
complex conjugate:
\be
\chi^*(\cdots,z_i,z_{i+1},\cdots)=e^{-i\pi\kappa\epsilon(z_i-z_{i+1})}
\chi^*(\cdots,z_{i+1},z_i,\cdots)\, , \ee
we reduce Eq.~\ref{intff} to the final expression for the form
factor
\be F_{3,2}(x)=\sqrt{3}\int d^2z\ \chi^*_3(z_1,z_2,x)\chi_2 (z_1,
z_2) \, . \label{m9} \ee

The calculations leading to Eq.~(\ref{m9}) can be generalized to
arbitrary $N$:
\be F_{N+1,N}(x)=\la \Psi_{N+1}|\fad(x)|\Psi_{N}\ra=\sqrt{N+1}\int
d^Nz\ \chi^*_{N+1}(z_1,\cdots,z_N,x)\chi_N(z_1,\cdots,z_N)\, .
\label{m11} \ee
This result follows from Eq.~(\ref{m10}) by noticing that the
statistical phase factors in the commutation relations
(\ref{com1})--(\ref{com3}) of the field operators are compensated by
the exchange property (\ref{m8}) of the wavefunctions. This means
that the pairing of the $\fad(x)$ operator with any of the
$\fa(y_j)$ operators produces $N+1$ identical terms in which the
coordinate $x$ is made the last coordinate of the wavefunction
$\chi_{N+1}$. After that, the integrals over $z$'s and remaining
$y$'s can be limited to the ordered regions $z_1>z_2> ... >z_N$ and
$y_1>y_2> ...>y_N$ giving directly (\ref{m11}).


\section{Thermodynamic Limit of Singular Sums}\label{TL}

In this appendix, we study the behavior of the functions defined by
Eqs.~(\ref{defg}), (\ref{defe}), and (\ref{defte}) in the
thermodynamic limit of large length $L$ of normalization interval.
We start with (\ref{defg}). In this case, the function summed over
the momenta $\lambda$ is sufficiently smooth, so that the anyonic
shift $2\pi\delta'/L$ of the momenta becomes negligible when $L
\rightarrow \infty$, and one can pass directly from the sum to the
integral over $\lambda$:
\be G(t,x)\equiv \lim_{L\rightarrow\infty}G_L(t,x)= \frac{2\pi}{L}
\sum_{ \lam_j\in \frac{2\pi}{L}(\mathbb{Z}+\delta')}e(\lam_j|t,x) =
\int_{-\infty}^\infty e(\lam|t,x)\ d\lam\, . \ee
The regularization $t\rightarrow t+i0$ for $e(\lam|t,x) =\exp(it \lam^2
-ix\lam)$ is implied in these expressions.

Next, we turn to Eq.~(\ref{defe}). In this case, the function under
the sum is no longer smooth in the thermodynamic limit. We transform
it by separating the singular part that can be summed explicitly:
\begin{eqnarray}
E(\mu_k|t,x) \equiv \lim_{L\rightarrow \infty} E_L(\mu_k|t,x) &=&
\frac{2\pi}{L}\sum_{\lam_j\in \frac{2\pi}{L} (\mathbb{Z}+\delta')}
\frac{e(\lam_j|t,x)}{\lam_j-\mu_k} \nonumber\\ &=&\frac{ 2\pi}{L}
\sum_{\lam_j\in \frac{2\pi}{L}(\mathbb{Z}+\delta')} \frac{e( \lam_j|
t,x)-e(\mu_k|t,x)}{\lam_j-\mu_k}+e(\mu_k|t,x)\sum_{n=-\infty}^\infty
\left(n-\frac{\kappa+1}{2}\right)^{-1}\, .\label{s1int} \end{eqnarray}
In the last line here we have used Eq.~(\ref{nonvanish}). The first
term in (\ref{s1int}) is now a smooth function, so as before, we can
directly replace the sum with the integral, since the anyonic shift
of the momenta does not affect the value of the integral. The
integral can then be transformed as follows:
\begin{eqnarray}
\int_{-\infty}^\infty d\lam\ \frac{e(\lam|t,x)-e(\mu_k|t,x)}{ \lam-
\mu_k} &=&\mbox{P.V.} \int_{-\infty}^\infty d\lam\ \frac{e(\lam|
t,x) }{ \lam-\mu_k}-e(\mu_k|t,x)\ \mbox{P.V.} \int_{-\infty}^\infty
\frac{d\lam}{\lam-\mu_k} \nonumber \\ &=&\mbox{P.V.}
\int_{-\infty}^\infty d\lam\ \frac{e(\lam|t,x)}{\lam-\mu_k}\,
.\label{s2int}
\end{eqnarray}
Under the natural
interpretation of the sum in the second term in (\ref{s1int}), it
can be simplified using formula 1.421.(3) of \cite{GR}, $\pi
\cot(\pi x)=(1/x)+2x\sum_{n=1}^\infty (x^2-n^2)^{-1}$:
\be \label{b6} \sum_{n=-\infty}^\infty\left(n-\frac{\kappa+1}{2}\right)^{-1} =
\pi \tan\left(\frac{\pi\kappa}{2}\right)\, . \ee
Collecting the two terms we  finally get
\be E(\mu_k|t,x)=\mbox{P.V.} \int_{-\infty}^\infty d\lam\
\frac{e(\lam|t,x)}{\lam-\mu_k}+e(\mu_k|t,x)\pi\tan\left(\frac{
\pi\kappa}{2}\right)\, . \ee

The function defined by Eq.~(\ref{defte}) is more singular than
$E(\mu_k|t,x)$ ((\ref{s1int}). To transform it, we use the same
strategy of separating the most divergent terms that can be summed
explicitly:
\begin{eqnarray}\label{b8}
\tilde E(\mu_k|t,x)&\equiv& \lim_{L\rightarrow\infty}\tilde E_L(
\mu_k|t,x)=\frac{4}{L^2}\cos^2(\pi\kappa/2)\sum_{\lam_j\in \frac{2
\pi}{L} (\mathbb{Z}+\delta')} \frac{e(\lam_j|t,x) }{ (\lam_j-
\mu_k)^2}\, ,\nonumber\\ & =&\frac{4}{L^2}\cos^2(\pi\kappa/2)
\left(\sum_{\lam_j\in \frac{2\pi}{L}(\mathbb{Z}+\delta')}
\frac{e(\lam_j|t,x)-e(\mu_k|t,x)}{(\lam_j-\mu_k)^2}+e(\mu_k|t,x)
\frac{L^2}{4\pi^2}\sum_{n=-\infty}^\infty\frac{1}{\left(n-\frac{
\kappa+1}{2}\right)^2}\right)\, . \end{eqnarray}
Defining
\be f(\mu_k)=\sum_{\lam_j\in \frac{2\pi}{L}(\mathbb{Z}+\delta')}
\frac{e(\lam_j|t,x)-e(\mu_k|t,x)}{\lam_j-\mu_k}\, , \ee
one has
\be \sum_{\lam_j\in \frac{2\pi}{L}(\mathbb{Z}+\delta')}
\frac{e(\lam_j|t,x)-e(\mu_k|t,x)}{(\lam_j-\mu_k)^2}= \frac{\partial
f(\mu_k)}{\partial \mu_k}+\frac{\partial e(\mu_k|t,x)}{\partial
\mu_k} \sum_{\lam_j\in
\frac{2\pi}{L}(\mathbb{Z}+\delta')}\frac{1}{\lam_j-\mu_k} \ee
Taking the limit $L\rightarrow\infty$ and using (\ref{s2int}) and
(\ref{b6}) in this equation we obtain
\be \lim_{L\rightarrow\infty} \frac{L}{2\pi} \sum_{\lam_j\in
\frac{2\pi}{L}( \mathbb{Z} +\delta')} \frac{e(\lam_j|t,x) -e(\mu_k|
t,x) }{(\lam_j-\mu_k)^2}= \frac{\partial}{\partial \mu_k}
\left(\mbox{P.V.} \int_{-\infty}^\infty d\lam\
\frac{e(\lam|t,x)}{\lam-\mu_k}\right) +\frac{\partial
e(\mu_k|t,x)}{\partial
\mu_k}\pi\tan\left(\frac{\pi\kappa}{2}\right)\, . \ee For the second
term in the R.H.S. of (\ref{b8}) we use the formula 1.422.(4) of
\cite{GR} $\pi^2/\sin^2(\pi x) =\sum_{n=-\infty}^\infty (n-x)^{-2}$
to get
\be \sum_{n=-\infty}^\infty \left(n-\frac{\kappa+1}{2} \right)^{-2} =
\frac{\pi^2}{\cos^2(\pi\kappa/2)}\, . \ee
Collecting all the terms we have the final result
\be \tilde E(\mu_k|t,x)=e(\mu_k|t,x)+\frac{2\cos^2(\pi\kappa/2)}{\pi
L}\frac{\partial e(\mu_k|t,x)}{\partial \mu_k}\pi\tan \left( \frac{
\pi\kappa}{2}\right) +\frac{2\cos^2(\pi\kappa/2)}{\pi L}\frac{
\partial }{\partial\mu_k}\left(\mbox{P.V.} \int_{-\infty}^\infty
d\lam\ \frac{e(\lam|t,x)}{\lam-\mu_k}\right)\, . \ee


\end{document}